# Beyond survivorship bias: REGARD for safety models in generative drug discovery

Andrew Dunn[1] and Adriana S. Beltran[2*]

[1]School of Mathematics, Georgia Institute of Technology, Atlanta, GA, USA
[2]Department of Genetics, University of North Carolina at Chapel Hill, Chapel Hill, NC, USA

[*]Corresponding author: Adriana S. Beltran, PhD, Department of Genetics, University of North Carolina at Chapel Hill, 120 Mason Farm Road, Genetic Medicine Building, Campus Box 7365, Chapel Hill, NC 27599, USA; beltran@med.unc.edu

## Abstract

Safety models used as new approach methodologies (NAMs) in drug discovery rely on evidence shaped by historical compound selection. However, compounds excluded for safety reasons left little or no outcome data. As generative AI brings similar compounds back into discovery, models are asked to assess risks for which the evidence is limited. Changes in the discovery workflow may also remove safeguards that served a biological safety function. Assessing whether existing evidence supports the changed workflow requires distinguishing three risks: the risk a model was evaluated against, the risk across all historical proposals, and the risk in its intended use. The difference between the first and last reflects shifts in context, generation, and selection. When safeguards precluded all outcome ascertainment, positivity fails, and reweighting cannot recover what was never measured. Thus, model performance may remain unchanged even when the evidence no longer justifies its use. We define guardrail displacement as the material weakening, removal, or bypass of historical safeguards and propose REGARD (Reassessing Evidence, Guardrails, and Authority for Responsible Discovery) to assess its consequences. REGARD assesses if historical selection has materially depleted outcome evidence and whether an empirically supported biological safety function has been lost without adequate replacement. Drug-induced liver injury and cardiac proarrhythmia illustrate both triggers and their remedies. REGARD determines if existing evidence suffices, challenge evidence is required, or model authority should be constrained. Justified reliance on a safety model therefore depends not only on its predictive performance but on the decision system in which it operates.

### Plain language summary

Safety models learn from the test results of compounds that organizations chose to synthesize. Compounds set aside over safety concerns left no such records. Generative AI can now bring similar molecules back into discovery, where the absence of recorded failures is a blind spot, not a safe zone. The model has not changed, but what it must assess has. Here, we propose REGARD, a framework that asks if past decisions left gaps in the evidence and whether removing a safeguard would leave a biological safety function uncovered. Answering these questions helps determine which rules can be retired and which safety functions must still be covered. If human-relevant methods, such as new approach methodologies, can provide that coverage, they can replace animal studies. Measuring the resulting change in animal use makes the contribution to the 3Rs explicit.

## 1 Introduction[1]

The current debate over new approach methodologies (NAMs) in drug development is framed as a trade-off between accelerating human-relevant research under the 3Rs and maintaining confidence that the established safety functions remain covered. However, this debate overlooks another problem: how much authority should we give to a safety model when artificial intelligence (AI) generates small-molecule candidates previously excluded from testing, leaving behind little or no evidence about their safety? This question is becoming more important as AI is reshaping drug discovery and as NAMs become more widely used in safety assessment. Importantly, the use of NAMs, including AI-based computational models, to reduce animal testing is encouraged by the US Food and Drug Administration (FDA) Roadmap to Reducing Animal Testing in Preclinical Safety Studies (FDA, 2025b, 2026a). One example of AI-driven drug discovery reaching clinical testing is rentosertib, a computationally generated molecule evaluated in a randomized phase 2a trial (Xu et al., 2025). Although the FDA has issued additional draft guidance emphasizing context of use, human biological relevance, and fitness for purpose, drug discovery remains outside its scope (FDA, 2026b).

We can address this question by looking at how the evidence was generated. Safety models are trained on compounds selected at different stages of drug discovery, before definitive outcomes could be measured. Selection may reflect structural alerts, reactive metabolites, mechanistic concerns, and expert judgment (Kutchukian et al., 2012; Stepan et al., 2011; Waring et al., 2015; Kalgutkar, 2020). This process has two consequences. First, it limits the evidence available for evaluating previously excluded compounds, and second, it may have served as a safety filter that the model was never designed to replace. Therefore, the model may learn relationships between chemical structures and measured safety outcomes without learning the reasons why some compounds were never tested.

Generative AI makes these concerns more critical because it can systematically explore previously avoided chemical regions and can integrate safety predictions directly into candidate optimization. A generator can optimize predicted safety without necessarily improving biological safety (Fannjiang et al., 2022; Langevin et al., 2023; Yoshizawa et al., 2025). At the same time, the fact that a chemical structure was previously avoided does not necessarily mean it is unsafe. Exclusion could have been based on scientific knowledge, technological limitations, or the risk tolerance prevailing at the time (Boike et al., 2022). Established concepts such as selective labels, missing-not-at-random outcomes, positivity, dataset shift, performative prediction, and applicability-domain analysis address different aspects of this problem (Lakkaraju et al., 2017; Petersen et al., 2012; Quiñonero-Candela et al., 2008; Perdomo et al., 2020). Newer work benchmarks the safety hazards of generated molecules (Xu et al., 2026), evaluates generator outputs through staged filters (Ryabchenko et al., 2026), and adds human-review checkpoints to redesign workflows (Kim and Hong, 2026). However, what these methods do not answer is what justifies carrying over the evidentiary authority of an existing safety model when the discovery workflow changes. Being able to make a prediction does not mean there is enough evidence to rely on it.

[1] **Abbreviations:** AI, artificial intelligence; ASME, American Society of Mechanical Engineers; BSEP, bile salt export pump; CiPA, Comprehensive in Vitro Proarrhythmia Assay; DILI, drug-induced liver injury; FDA, US Food and Drug Administration; hERG, human ether-à-go-go-related gene; ICCVAM, Interagency Coordinating Committee on the Validation of Alternative Methods; ICH, International Council for Harmonisation of Technical Requirements for Pharmaceuticals for Human Use; NAM, new approach methodology; OECD, Organisation for Economic Co-operation and Development; 3Rs, replacement, reduction, and refinement

In this paper, we introduce guardrail displacement, which refers to cases where a historical safety-motivated selection practice is materially weakened, removed, or bypassed when a workflow re-enters the chemical region it affected. We propose REGARD (Reassessing Evidence, Guardrails, and Authority for Responsible Discovery) for safety assessment. REGARD is a provenance-conditioned qualification framework with two complementary non-nested triggers. The evidence-support trigger addresses depletion of relevant outcome evidence for the newly reachable region, whereas the functional-replacement trigger addresses the loss of a credible biological safety function without adequate replacement. Both require material re-entry or consequential decision exposure, together with consequential model authority. We do not assume that historical rules are correct; scientific advances may justify dropping them or replacing them with more effective human-relevant methods. Our objective is to determine when additional evidence, safeguards, or limits on model authority are needed, while reducing unnecessary animal testing without preserving outdated safety rules.

## 2 How historical selection influences available safety evidence

To understand how historical selection shapes the available evidence, we need to distinguish changes in the distribution of candidates, in the biological context, and in the availability of safety data. Each change introduces a different source of uncertainty that affects which outcomes are available and how well the resulting evidence supports the model predictions.

### 2.1 Selection and limits of statistical correction

The compounds under evaluation may differ from those used to train a model while the relationship between molecular features and safety outcomes remains stable. In contrast, changes in exposure, formulation or biological context could alter this relationship independently (Quiñonero-Candela et al., 2008). An additional concern is selective labels, in which outcomes are observed only for the compounds selected for testing (Lakkaraju et al., 2017). Statistical correction relies on three assumptions (Cole and Hernán, 2008; Petersen et al., 2012): that a sufficient set of variables influencing selection was measured; that once these are controlled for, selection does not depend on the outcome; and that observation has positive probability across the population whose mean is sought (overlap). Weighting also requires known or consistently estimated observation probabilities. These conditions may not be met, particularly if compounds were discarded because of mechanistic concerns or expert judgment that was never recorded.

### 2.2 A stage-resolved formulation

The mismatch must be located, and we should define which expectation a safety model has been evaluated against. To make this precise, let $X$ denote the molecular features of a candidate, $C$ the context of use, $Y$ the outcome relevant to that use, and $m$ the safety model. Let $L(m(X), Y)$ be the loss incurred when $m$ is applied to $X$ and the outcome $Y$ is realized. Let $S_1, \dots, S_K$ indicate passage through the successive stages of proposal, synthesis, screening, advancement and definitive evaluation, with $S_k = 1$ implying $S_{k-1} = 1$ for $k = 2, \dots, K$. Let $O$ be a global observation indicator, so that $O = 1$ when a candidate reaches definitive evaluation and its result enters the accessible evidence base, and $O = 0$ otherwise. Here, definitive means sufficiently informative for the use claim, and not necessarily a human outcome. Let $P_H$ and $P_G$ denote the historical and the intended proposal-stage candidate distributions, $P^*(Y \mid X, C)$ the outcome-generating law in context $C$. Let $Q_H$ be the full historical joint law over $X, Y, S_1, \dots, S_K$ and $O$ which these induce together with the stage-specific selection and reporting mechanisms. A typical evidence-generating path is $X \to S_1 \to S_2 \to \cdots \to S_K \to O$, and model development is conditioned on $O = 1$ rather than on all of the candidates which could have been considered (Fig. 1). Evaluation therefore measures one expectation while the intended use requires another, and a third is needed to tell them apart. Thus, three risks must be distinguished: the risk on the evidence observed (Equation 1), the risk under the historical proposal distribution without conditioning on the downstream selection (Equation 2), and the risk in the intended use (Equation 3).

$$R_{\mathrm{obs}}(m) = \mathbb{E}_{Q_H}[L(m(X), Y) \mid O = 1] \quad \text{(Eq. 1)}$$

$$R_H(m) = \mathop{\mathbb{E}}_{\substack{X \sim P_H \\ Y \sim P^*(\cdot \mid X, C_H)}} [L(m(X), Y)] \quad \text{(Eq. 2)}$$

$$R_G(m) = \mathop{\mathbb{E}}_{\substack{X \sim P_G \\ Y \sim P^*(\cdot \mid X, C_G)}} [L(m(X), Y)] \quad \text{(Eq. 3)}$$

To separate a change in the biological or exposure context from a change in the proposal distribution, we define the cross-setting risk $R_{G,H}(m)$ by applying the intended proposal distribution $P_G$ under the historical outcome context $C_H$:

$$R_{G,H}(m) = \mathop{\mathbb{E}}_{\substack{X \sim P_G \\ Y \sim P^*(\cdot \mid X, C_H)}} [L(m(X), Y)]$$

The total difference is what we call the transfer problem, and it is represented as:

$$R_G(m) - R_{\mathrm{obs}}(m) = \left[R_G(m) - R_{G,H}(m)\right] + \left[R_{G,H}(m) - R_H(m)\right] + \left[R_H(m) - R_{\mathrm{obs}}(m)\right] \quad \text{(Eq. 4)}$$

The three terms in Equation 4 represent, respectively, a context or mechanism shift, a generator or candidate-distribution shift, and the historical selection or observation contrast.

Equation 4 is an accounting identity. By itself, it does not identify any of its components from the selected archive without additional assumptions or evidence. Of these, REGARD addresses the selection term $R_H(m) - R_{\mathrm{obs}}(m)$, to include all selection and reporting stages, so that attribution to a safety guardrail will require separate evidence. Attribution between the context and generator shifts is order-dependent. Introducing the historical proposal distribution under the target context gives an equally valid decomposition, which allocates those two differently while leaving the total contrast unchanged.

If a relevant component had zero chance of an accessible outcome, this means that the outcome cannot be recovered from historical data without making additional assumptions or bringing outside evidence. Very small selection probabilities can make estimates unstable even when identification is possible (Cole and Hernán, 2008; Petersen et al., 2012). Thus, partial identification may be preferable to poorly supported estimates (Manski, 2003). Existing methods address statistically correctable selection effects (Cole and Hernán, 2008; Liu and Kashima, 2022), but where a consistently applied guardrail prevents ascertainment through every available route, no reweighting recovers the missing quantity (see Equation S1-3 in Supplementary Methods S1).

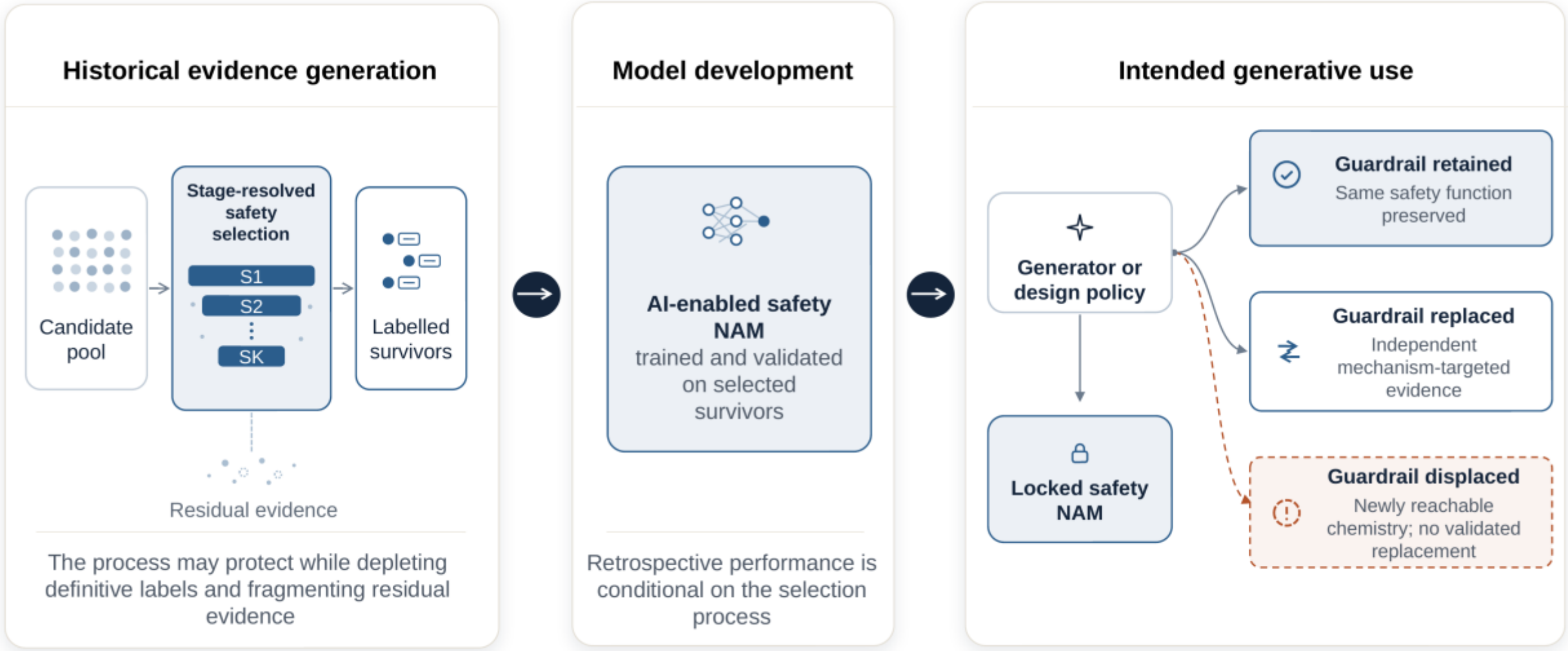

**Fig. 1: Historical safety selection can both condition outcome availability and perform a risk-control function.** Stage-resolved selection removes candidates before their outcomes are measured, so the archive retains definitive labels for the survivors whose results are accessible ($O = 1$) and only fragmented evidence for the rest. A model trained and validated on those survivors does not automatically reproduce either role. When a generative workflow re-enters the affected chemistry, the historical guardrail may be retained, independently replaced, or displaced without an adequate replacement. Independent replacement requires empirical evidence of adequate coverage of the relevant safety function.

## 2.3 Applicability domains and evidentiary support

Applicability-domain analysis remains essential to evaluate if a safety prediction is supported by the available evidence. The Organisation for Economic Co-operation and Development (OECD) provides principles for defining and assessing the applicability domain of quantitative structure–activity relationship models, although mechanistic domains can extend beyond chemical descriptor similarity (OECD, 2014, 2023; Schultz et al., 2007; Horvath et al., 2009). Conformal prediction provides prediction sets for individual compounds whose coverage is guaranteed on average over repeated draws, under specified assumptions of exchangeability or shift-adjustment (Norinder et al., 2014; Alvarsson et al., 2021; Fannjiang et al., 2022).

Exclusion provenance adds value when it provides decision-relevant information beyond chemical similarity, mechanistic reasoning, uncertainty estimation and selection-bias correction. Two compounds may have similar chemical distances and uncertainty scores, yet one may come from an underexplored region and the other from a region repeatedly abandoned after documented safety signals. Marginal calibration or conformal coverage does not guarantee comparable performance in selectively under-observed subgroups, nor does it establish functional replacement. Shift-aware conformal methods address specified feedback mechanisms under their own assumptions (Fannjiang et al., 2022). Where established controls demonstrate sufficient evidence and retention or replacement of the relevant function, no additional provenance-specific requirement is warranted.

Historical selection can both limit the available outcome evidence and serve a safety function. In order to distinguish between these two effects, it is first necessary to know why the candidates failed. The reasons for failure are multiple and include safety concerns, failure of the synthesis, insufficient potency, commercial decisions and incomplete reporting. Fragment prioritization reflects chemists' choices (Kutchukian et al., 2012), nominated drug candidates are terminated for safety, efficacy or commercial reasons (Waring et al., 2015), and failed syntheses can preclude testing, as illustrated in materials research (Raccuglia et al., 2016). Moreover, early safety findings may inform the progression of a candidate even when the later outcomes are not available (Bowes et al., 2012; Sanz et al., 2017; Shinozawa et al., 2025). Thus, the gaps in the evidence cannot be attributed to safety selection without examining the underlying workflow. We propose a stage-resolved audit to trace eligible candidates from the relevant selection decision to the availability of downstream outcomes. Our audit links dated safety rules or findings to the advancement or the termination of the candidate, records the number of candidates at each stage and distinguishes the safety-related exclusions from competing reasons. The audit separates the outcomes that were never generated from those of studies that were completed but remain inaccessible. Termination records, in turn, document overrides and changes in the safety rules that can strengthen this reconstruction. However, these records cannot independently establish what would have happened in the absence of the guardrail. Evidence that the selection prevented the outcome from being measured does not establish that the excluded compounds were toxic. For example, a candidate terminated after a positive reactive-metabolite finding may present informative early safety evidence, but no subsequent outcome data. In contrast, a candidate abandoned because the synthesis failed, or one for which the studies were completed but remain inaccessible, represents a different situation. Together, these distinctions determine what the archive can support and where additional evidence is required.

## 3 Operational triggers and materiality

The two consequences of selection lead to different questions: first, what the historical evidence can establish, and second, what safety protection may be needed to replace the old selection process. A replacement assay may address the second question without establishing the first, whereas additional outcome evidence does not demonstrate that a displaced function remains covered (see Tab. 1).

### 3.1 Reconstructing guardrail domains

For each historical guardrail, we document the period, rule version, inputs, thresholds, overrides and classification reliability, evaluating exposure-dependent rules under prespecified conditions.

Inferred domain membership must not be treated as proof that a specific molecule was historically excluded. We retain overlapping guardrails and competing exclusion reasons.

**Tab. 1: Key concepts for provenance-conditioned qualification**

| Concept | Definition |
|---|---|
| Guardrail domain | A time-, sponsor-, rule- and context-specific chemical or mechanistic region to which a historical selection practice applied, denoted $A_g$. Membership may be deterministic or probabilistic, and a candidate may belong to several overlapping domains. |
| Exclusion provenance | The documented history of why context-relevant outcomes were never measured or their results remained inaccessible, including safety and non-safety reasons, uncertainty, rule versions, overrides and multiple contributing causes. For a newly generated candidate, provenance is inferred from its relation to a reconstructed domain; it is not a claim that the molecule was historically considered. |
| Evidence-support trigger | A trigger for additional assessment of evidentiary support when historical safety selection materially depleted context-relevant definitive outcomes, the intended use materially re-enters the region or creates consequential decision exposure within it, and the model will carry consequential authority (Equation 13). |
| Functional-replacement trigger | A trigger requiring retention or independent replacement of a biologically plausible and empirically supported risk-control function when the intended use materially re-enters the region or creates consequential decision exposure within it, the model will carry consequential authority, and no adequate retention or replacement is in place (Equation 14). |
| Guardrail displacement | Material weakening, removal or bypass of a historical safety-motivated selection practice in a workflow that materially re-enters the affected region. Whether a credible safety function is left without adequate retention or replacement is assessed separately (Equation 14). |
| Provenance-conditioned qualification | A qualification or credibility process in which the permitted model authority depends both on performance and on how the supporting evidence and the future inputs were generated. |
| REGARD | Reassessing Evidence, Guardrails, and Authority for Responsible Discovery: the provenance-conditioned qualification framework proposed here, comprising the two triggers of Equations 13 and 14, the materiality quantities of Equations 5–12, and the six review steps of Section 6. |

To formalize this uncertainty, we define guardrail domains in their historical and decision-making contexts. We write $A_g$ for the domain of guardrail $g$, that is, the chemical or mechanistic region, specific to a time period, a sponsor, a rule version and a context, to which the historical selection practice applied. Membership in $A_g$ may be deterministic or probabilistic, and a candidate may belong to several overlapping domains. A structural rule may define $A_g$ directly. An expert judgment is instead reconstructed as an explicit or probabilistic classifier, whose sensitivity $\mathrm{se}$ and specificity $\mathrm{sp}$ are reported together with the inter-rater agreement wherever these are estimable. The resulting classification error is propagated into the reachability and trigger estimates described in the following section (see also Supplementary Methods S3).

### 3.2 Reachability and consequential decision exposure

To assess the materiality of a change in a discovery workflow, we distinguish two quantities: reachability at the proposal stage, which compares the proportions of candidates within a guardrail domain under the historical and intended workflows, and exposure to consequential model-governed decisions. When proposal-stage records are unavailable, the survivor archive reflects earlier selection rather than proposal-stage prevalence. The following bounds distinguish what can and cannot be established from that archive.

Formally, let $p_G = P_G(X \in A_g)$ and $p_H = P_H(X \in A_g)$ denote the intended and the historical proposal-stage probabilities of membership in the guardrail domain. The absolute reachability contrast is then:

$$\Delta_g = p_G - p_H \qquad \text{(Eq. 5)}$$

Equation 5 is identifiable where adequate proposal-stage records exist, but it is not identifiable from a survivor archive alone. Let $B = 1$ denote inclusion in the relevant historical archive, with $B = S_K$ when all definitively evaluated candidates are available and $B = O$ when only those results which entered the accessible evidence base are available, and let $p_{H,B} = Q_H(X \in A_g \mid B = 1)$ denote the prevalence of the guardrail domain in that archive. An archive-based upper bound requires the monotone-depletion assumption, that is, candidates inside the domain were no more likely to enter the archive than candidates outside it:

$$Q_H(B = 1 \mid X \in A_g) \leq Q_H(B = 1 \mid X \notin A_g) \qquad \text{(Eq. 6)}$$

Under Equation 6, $p_{H,B} \leq p_H$, and:

$$\Delta_g \leq \overline{\Delta}_g \coloneqq p_G - p_{H,B} \qquad \text{(Eq. 7)}$$

Equation 8 gives an archive-based sample estimator of the upper-bound quantity:

$$\widehat{\overline{\Delta}}_g = \hat{p}_G - \hat{p}_{H,B} \qquad \text{(Eq. 8)}$$

The bound is one-sided in the relevant direction. A prespecified increase in the proposal rate can be ruled out only where a one-sided upper confidence limit for the bound of Equation 7 lies below the materiality threshold and the monotone-depletion assumption is defensible. A large bound cannot establish displacement. Point identification requires proposal-stage records that supply a denominator, such as complete ideation logs or screened virtual libraries; termination rationales recorded at the guardrail stage add the reasons for exclusion but not the denominator.

The true proposal-stage relative re-entry ratio, Equation 9, is:

$$\rho_g = \frac{p_G}{p_H}, \quad p_H > 0 \qquad \text{(Eq. 9)}$$

and under the same monotone-depletion assumption of Equation 6 the survivor archive supports only the upper-bound proxy of Equation 10:

$$\rho_g \le \overline{\rho}_g := \frac{p_G}{p_{H,B}}, \quad p_{H,B} > 0 \qquad \text{(Eq. 10)}$$

estimated by:

$$\widehat{\overline{\rho}}_g = \frac{\hat{p}_G}{\hat{p}_{H,B}} \qquad \text{(Eq. 11)}$$

The quantity in Equation 11 is what we call survivor-relative enrichment, and it must not be reported as the true proposal-stage re-entry ratio. Where the observed historical archive count is zero or near zero, counts and interval bounds must be reported instead of a smoothed, effectively infinite ratio; when the count is zero, no empirical ratio is reported. The confidence procedures for Equations 8 and 11, and the correction for classification error in a reconstructed domain, are given in Supplementary Methods S2. Under non-differential misclassification by a classifier with sensitivity $se$ and specificity $sp$, the apparent absolute bound is attenuated toward zero by the factor $se + sp - 1$ when $0 < se + sp - 1 < 1$. With imperfect specificity, the apparent enrichment ratio is also pulled toward one when the relevant denominators are positive. For positive absolute bounds and a positive materiality threshold, uncorrected misclassification can render the rule-out test anti-conservative; uncertainty in the correction must therefore be propagated.

The second quantity measures how many candidates within the guardrail domain reach a decision at which model outputs influence advancement or rejection. Candidates may be filtered out before that point, and decision counts depend on workflow throughput and progression as well as proposal prevalence. Materiality therefore depends on absolute target exposure, decision frequency and intended model authority.

Let $V_j = 1$ indicate that a candidate reaches a consequential model-governed decision under process $j \in \{H, G\}$, and let $N_j$ be the number of candidates entering that workflow over a prespecified period or budget, with probabilities taken under $Q_j$, the joint law of candidates and workflow indicators under process $j$. For fixed throughput, the expected decision counts and their contrast are:

$$\mu_{g,j} = N_j Q_j\big(X \in A_g, V_j = 1\big), \quad \Delta_g^{\text{dec}} = \mu_{g,G} - \mu_{g,H} \qquad \text{(Eq. 12)}$$

Equation 12 prevents proposal-stage proportions from being treated as sufficient evidence of consequential decision exposure. Its identification requirements and per-entrant counterpart are given in Supplementary Methods S4. None of the reachability, enrichment or decision-frequency measures is by itself a qualification trigger. When throughput is random and associated with candidate composition or progression, the expected count is averaged over their joint distribution as described in Supplementary Methods S4.

### 3.3 Evidence-support trigger

The evidence-support trigger focuses on the adequacy of the outcome evidence (Fig. 2). We define three indicators: $D_g = 1$ when the historical safety-motivated selection materially depleted the context-relevant definitive outcomes within $A_g$; $E_g = 1$ when the target workflow materially re-enters $A_g$ or creates consequential decision exposure within it, in the sense of Equations 5 and 12; and $U_g = 1$ when the model carries consequential authority over the affected candidates. For resolved components, the trigger is the conjunction:

$$T_g^{(\mathrm{E})} = \mathbf{1}\{D_g = 1,\ E_g = 1,\ U_g = 1\} \qquad \text{(Eq. 13)}$$

Depletion must be demonstrated using stage-resolved evidence. Chemical sparsity or low prevalence in an archive is not enough, because either can arise for reasons unrelated to safety. When the trigger applies, model qualification requires an explicit assessment of the support available in the newly reachable region. Several responses are possible at that point: targeted human-relevant challenge evidence, independent corroboration, abstention, referral, or limits on the model's authority. A comment on the conjunction logic applies to both triggers. When any required condition is known to fail, the conjunction is zero. When all of them are established, it is one. In every other case the result remains unknown, and the unresolved components are documented even when another condition already determines the value of the conjunction. A trigger value of zero means that this particular additional qualification requirement has not been established and it does not establish that the compound is safe. Unknown provenance, in the same way, implies neither hazard nor safety.

### 3.4 Functional-replacement trigger

The REGARD functional-replacement trigger asks if a credible historical safety function remains adequately addressed, and this question is independent of whether the historical selection depleted the outcome evidence. This trigger shares $E_g$ and $U_g$ with the evidence-support trigger but does not require $D_g$. We define $F_g = 1$ when a biologically reasonable risk-control function is supported by relevant empirical evidence, interpreted with mechanistic knowledge and with documented expert judgment, and $Z_g = 1$ when that function is retained, or independently replaced under specified implementation conditions and thresholds, and $Z_g = 0$ when no such retention or replacement is in place or when one has been assessed and found inadequate. For these components the trigger is:

$$T_g^{(\mathrm{F})} = \mathbf{1}\{F_g = 1,\ E_g = 1,\ U_g = 1,\ Z_g = 0\} \qquad \text{(Eq. 14)}$$

Here $Z_g$ enters the conjunction with the value zero. The trigger fires when the function is credible, there is material re-entry or consequential decision exposure in the domain, the model carries consequential authority, and no adequate retention or replacement of the function is in place. If any required condition is known to fail, the trigger is zero. Where a replacement exists but its assessment is incomplete, or its coverage is unknown, $Z_g$ is recorded as unknown. The trigger is then zero if another required condition fails, and is otherwise recorded as unknown. When the trigger applies, historical model performance alone cannot justify unchanged high-authority use. The two triggers are complementary and non-nested: either may apply without the other, and no assumption of statistical independence between them is imposed. Where a historical rule is unsupported or redundant, it may be retired without an additional functional replacement once adjudicated; an evidence-support burden may nevertheless remain if Equation 13 is satisfied.

Overinclusive rules may be replaced by more discriminating methods. Our objective is to preserve, and where possible improve, the biological assessment of safety rather than reproduce historical exclusions. When a function or its replacement remains unresolved, we constrain reliance on the model according to the consequences of an error.

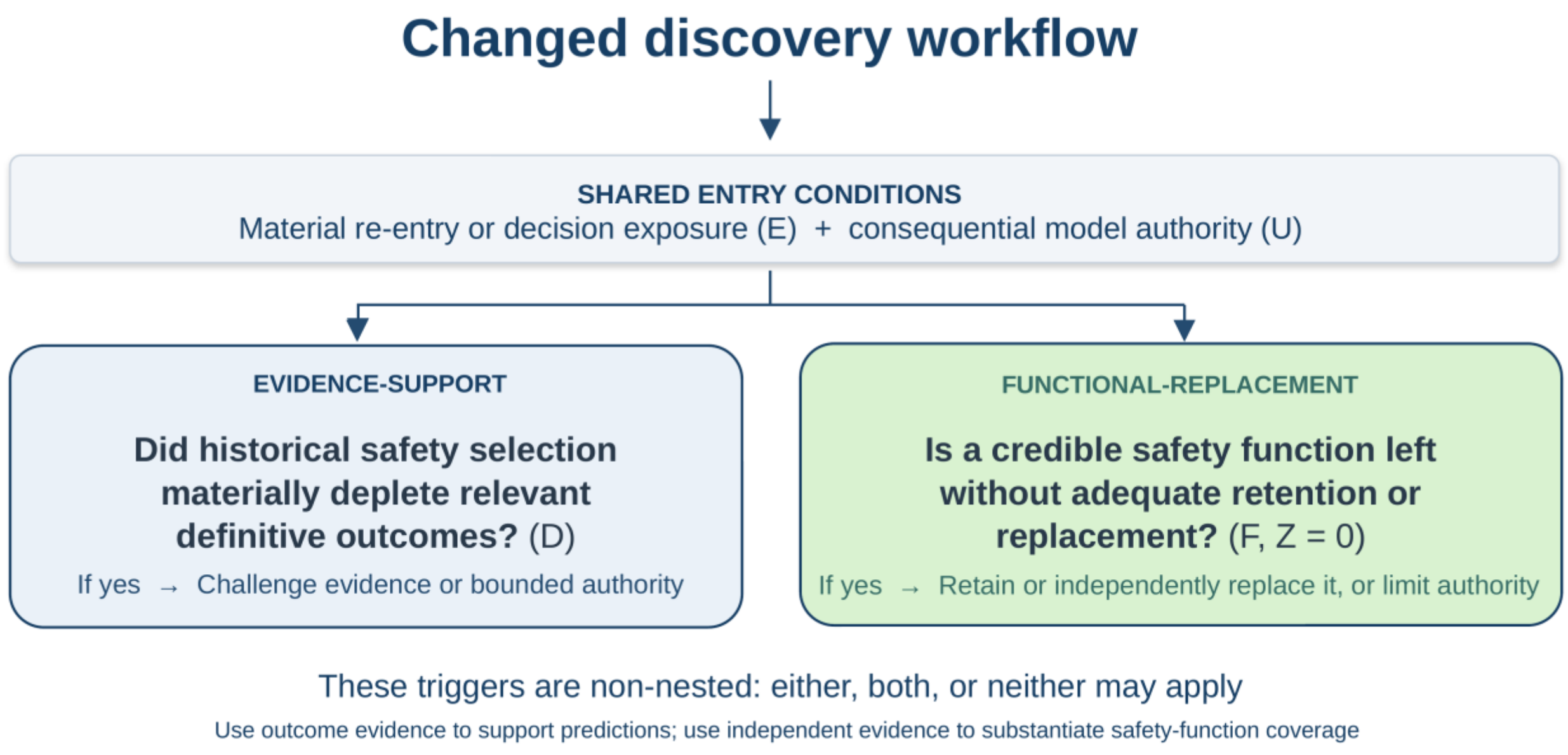

**Fig. 2: Two distinct REGARD qualification triggers.** Both triggers require material re-entry into the affected chemical region or consequential decision exposure (E), together with consequential model decision authority (U). The evidence-support trigger additionally requires material depletion of context-relevant definitive outcomes by historical safety selection (D). The functional-replacement trigger instead requires a biologically credible historical safety function (F) that is neither adequately retained nor replaced (Z = 0), and it does not require depletion. Either may apply, both may apply, or neither. Domain indices are omitted for readability; unresolved conditions are evaluated according to the three-valued convention in Section 3.3.

### 3.5 Independent evidence and functional replacement

Independent evidence is only useful when it can fail in different ways. We require that the evidence have materially different failure modes from those of the safety model and from those of the displaced guardrail. Independence is assessed across data sources, measurement principles, biological systems, endpoints, mechanistic assumptions, model architecture, laboratories, calibration data and decision governance, and any shared dependencies must be disclosed. The replacement is evaluated against the relevant biological risk and the decision consequences, not against agreement with the historical filters. We prespecify the relevant exposure conditions, decision thresholds, consequences of a severe false negative and rules for referral or abstention. Where appropriate, justified non-inferiority margins may be used against a satisfactory reference assessment. However, agreement with an inadequately supported historical rule does not suffice. Improved human-relevant methods may provide greater discrimination and a stronger assessment of safety without reproducing every historical positive.

### 3.6 Hypothetical example

Suppose that 8% of the historical proposals belonged to a reactive-metabolite guardrail domain, whereas among the compounds definitively evaluated only 1% belonged to it. A new generator now proposes 15% within the same domain. The actual increase at the proposal stage is 7 percentage points (15% − 8%), and the corresponding ratio is 1.875. A comparison with the survivor archive gives a different result, namely a difference of 14 percentage points (15% − 1%) and a survivor-relative enrichment of 15. Under monotone depletion that difference of 14 percentage points provides an upper-bound quantity. It is not an estimate of the actual increase at the proposal stage, and the survivor-relative enrichment must likewise not be read as the proposal-stage ratio (Supplementary Methods S2). We assess the two qualification triggers separately. The evidence-support trigger applies once historical depletion of the relevant outcome evidence, material re-entry or decision exposure, and consequential model authority have all been established. For the functional-replacement trigger the historical depletion plays no role: what matters is whether a credible safety function lacks adequate current coverage, under material exposure and consequential authority. Such coverage may be provided by a qualified liver system, used together with an exposure model and an orthogonal metabolite assay. Otherwise, the model's authority may require independent corroboration or referral.

## 4 Two safety domains illustrate REGARD boundaries

Two biological examples show why evidence support and functional replacement must be evaluated separately. Drug-induced liver injury shows what happens to the evidence when historical attrition has shaped it. Cardiac proarrhythmia, in contrast, is a case in which a historical rule was improved without the underlying safety objective being abandoned. Together they illustrate the two triggers of REGARD.

### 4.1 Drug-induced liver injury: limitations of historical safety evidence

Drug-induced liver injury (DILI) involves diverse mechanisms including reactive metabolites, mitochondrial dysfunction, the inhibition of the bile salt export pump (BSEP), lysosomal injury, immune effects and interactions with exposure and host factors (Mosedale and Watkins, 2017; Kullak-Ublick et al., 2017). A greater severity of DILI has been reported when mitochondrial and BSEP liabilities are combined (Aleo et al., 2014). Consistent with these mechanisms, structural alerts and mechanistic safety findings may prompt a redesign, additional testing or a termination, and they therefore define which candidates generate downstream outcomes. However, these signals do not independently establish human toxicity (Stepan et al., 2011; Kalgutkar, 2020).

Shinozawa et al. (2025) investigated TAK-994, an orexin receptor 2 agonist withdrawn from a phase II trial after severe DILI appeared. Neither the rat nor the nonhuman primate repeat-dose studies detected a hepatotoxic liability early; mouse studies conducted during clinical development later showed periportal single-cell hepatocellular necrosis at high doses. In hepatic cell culture, HepG2 cytotoxicity, mitochondrial toxicity and BSEP inhibition met the reduced-risk criteria at all clinical doses, whereas the margins in human liver microtissues and for hepatocyte bile transport fell below those criteria at the highest dose (180 mg twice daily). A covalent binding liability was uncovered subsequently, following the hepatic metabolism of the compound. The same work reports a retrospective analysis in which a HepG2 cytotoxicity assay and a human liver microtissue assay were applied to 51 small molecules terminated in a pharmaceutical pipeline because of direct or intrinsic liver toxicity. One or both assays identified a liability in 38 of these compounds, and neither detected one in 13. The overlap was only partial, with 18 compounds detected by both assays, 5 by HepG2 alone and 15 by the microtissue assay alone. Clinical DILI status was not established for the full cohort, although two of these compounds were terminated in phase III for hepatotoxicity. These findings illustrate the value of the historical attrition records and the limitations of the existing safety assays. However, they do not establish that a guardrail displacement caused a safety failure. The archive therefore records where assays disagree rather than which compounds would cause DILI in humans. Differences in DILI study designs and in control compounds limit how far these particular numbers generalize (Segovia-Zafra et al., 2024).

### 4.2 Cardiac proarrhythmia: replacing historical safety functions

Cardiac proarrhythmia is a different kind of case. Here the question is whether safety assessment preserves or improves a biological risk-control function, not if we can reproduce the historical exclusions. Inhibition of the human ether-à-go-go-related gene (hERG, KCNH2) channel and prolongation of the QT interval can both be overinclusive indicators of torsade de pointes risk, because other ion currents influence the net electrophysiological response. Verapamil is the familiar example, since it blocks both hERG and $Ca_V1.2$ channels and, although it prolongs the QTc interval (placebo-corrected 13.9 ms at peak concentration on day 3; 90% CI 10.0 to 17.8), it does not prolong the J–Tpeakc interval (−2.6 ms; 90% CI −11.4 to 6.1) and carries a low risk of torsade de pointes (Vicente et al., 2019). The Comprehensive in Vitro Proarrhythmia Assay (CiPA) integrates multichannel measurements, mechanistic modeling, human stem-cell-derived cardiomyocytes and clinical electrocardiographic evidence, and it thereby supports a more discriminating risk assessment (Colatsky et al., 2016; Vicente et al., 2019; FDA, 2026c). This does not render hERG testing obsolete. Nor does it establish that every multichannel system provides an adequate replacement. CardioGenAI incorporates predicted hERG, $Na_V1.5$ and $Ca_V1.2$ activities to guide molecular redesign, and its authors intend it for generating molecular hypotheses (Kyro et al., 2025). However, integrating these predictors into a generative design does not demonstrate that an earlier safety function has been replaced. Such a claim requires evidence addressing the relevant exposure, mechanisms, newly reachable chemistry and decision thresholds.

## 5 From model credibility to evidentiary authority

A reliable safety model does not by itself justify a consequential decision in a discovery workflow that has changed. The appropriate level of reliance depends on the influence of the model and the consequences of error, including unsafe advancement, rejection of useful candidates, delayed referral and unnecessary testing.

Evidentiary authority belongs to the complete decision system, and not to the model alone. This system comprises candidate generation, reachable chemistry, safety assays, independent controls, human oversight and management of change. A change in any component may alter justified reliance even when retrospective model performance remains unchanged. A model has high authority when its output becomes the sole or effectively decisive basis for consequential action, as can occur when an independent safety gate is removed.

### 5.1 Relationship to regulatory and credibility frameworks

Existing frameworks recognize that evidence requirements should reflect model influence and the consequences of error. The American Society of Mechanical Engineers (ASME) V&V 40 standard, FDA draft guidance on AI supporting regulatory decisions, and International Council for Harmonisation (ICH) M15 guideline provide relevant principles (ASME, 2018; FDA, 2025a; ICH, 2026). None addresses generative safety systems at the discovery stage: ASME V&V 40 addresses computational models of medical devices, whereas ICH M15 addresses model-informed drug development. The FDA's 2026 draft NAM guidance distinguishes validation from formal qualification and excludes discovery-stage use (FDA, 2026b). Substantial challenges to validation and acceptance of AI-based NAMs remain (Hartung and Kleinstreuer, 2025). REGARD extends existing context-of-use and credibility assessments by asking how historical safety selection affects the transfer of evidentiary authority, without creating a separate regulatory framework (Tab. 2).

### 5.2 Matching model authority to evidence

For low-influence use, such as exploratory prioritization followed by independent confirmation, the established assessments of context of use, applicability domain, uncertainty and credibility will generally suffice. At intermediate influence, corroboration should fail in ways that are materially independent of both the model and any displaced safeguard. Where predictions determine a consequential decision, the evidence-support trigger calls for targeted human-relevant challenge evidence, or for continued independent confirmation in newly reachable regions. The functional-replacement trigger instead asks for evidence that the relevant biological function remains adequately covered, whatever the historical outcome depletion. When neither trigger applies and existing controls support the intended use, REGARD requires nothing further. However, where uncertainty remains unresolved, the consequences of error limit how much authority the model may carry, and consulting regulators or another authorizing body early may help establish standards for the proposed high-authority use.

## 6 Implementing REGARD

REGARD implements provenance-conditioned qualification in six steps (Tab. 3), building on principles established for NAMs (ICCVAM, 2024; FDA, 2025a, 2026b; ICH, 2026; Kopańska and Hartung, 2026).

### 6.1 Define intended use and model authority

We specify the discovery stage, product scope, reachable chemistry, generator and optimization procedure, downstream assays, human oversight, permitted model-driven actions and consequences of error. We assess guardrail displacement separately from the credibility and current coverage of the historical function, recording its status as retained, independently replaced, unsupported, unresolved or materially unaddressed.

**Tab. 2: Qualification implications in different target evidence systems**

| Target evidence system | Historical selection or safety-function condition | Qualification implication |
|---|---|---|
| Conventional evaluator use | Candidate generation, selection, and safety screening remain materially similar to those in the process which generated the evidence. | The existing context-of-use, applicability-domain, uncertainty, and credibility evidence may support the claim, subject to ordinary change control. |
| Generative use: credible guardrail retained | The original safety rule, or a demonstrably equivalent safeguard, remains in effect. | It is necessary to verify the implementation and thresholds, together with the interactions with the generator, and to monitor changes in the target distribution and evidence system. Evidence support must be assessed separately. |
| Generative use: independent replacement | The historical rule is retired, but the credible safety function is covered by an orthogonal human-relevant assay, a mechanistic model, an ensemble, or a gate. | The replacement must be qualified for the mechanism, exposure, reachable chemistry, and decision consequence, and prediction support must be assessed separately. These are separate assessments and agreement with the old rule is not required. |
| Generative use: evidence gap without a credible function | Historical selection depleted definitive outcomes, and the adjudication found no material safety function that required retention or replacement. | The rule should not be preserved by inertia. The newly reachable evidence gap must be addressed with challenge data, abstention, referral, or bounded authority. |
| Generative use: credible function not replaced | The workflow materially re-enters the guardrail domain or creates consequential decision exposure there. The historical function is credible, but no proper replacement performs it. | Historical performance alone is insufficient to support sole-source or high-authority use, which requires mechanism-targeted challenge evidence or continued independent corroboration showing that a retained or replacement safeguard adequately covers the credible function. |
| Provenance undetermined | It is not possible to establish why the region lacked context-relevant definitive outcomes or if a credible safety function existed. | Neither hazard nor safety should be inferred. It is necessary to make the uncertainty explicit and to set reliance and prospective evidence requirements in proportion to decision consequences. |

### 6.2 Adjudicate the biological safety function

Historical exclusions are not evidence of toxicity. Signals from reactive metabolites, BSEP and ion channels must be read in light of exposure, metabolism, detoxification, competing mechanisms and physiological conditions (Stepan et al., 2011; Kalgutkar, 2020; Aleo et al., 2014; Colatsky et al., 2016; FDA, 2026c). We adjudicate their protective value and current coverage as described in Section 3.4. Exclusions not motivated by safety may still leave evidence gaps to address, but they do not on their own call for functional replacement.

**Tab. 3: Six linked questions for REGARD**

| Component | Established validation question | REGARD extension |
|---|---|---|
| 1. Context of use and authority | What question, endpoint, decision, product class and uncertainty does the NAM address? | Which historical stages determine outcome availability, what regions become newly reachable, and how much authority will the model carry? |
| 2. Biological relevance | Which human mechanisms and system features are captured or omitted? | What risk was the historical exclusion intended to control, and what evidence shows that the function was credible, weak, redundant or over-inclusive? |
| 3. Technical characterization | How reliable, reproducible, calibrated and bounded is the assay or model? | How were candidates selected and labeled, how accurately is the guardrail domain reconstructed, and what blinded challenge evidence tests the newly reachable region and any displaced function? |
| 4. Transparency and auditability | Can the model, data, assumptions, limitations and decisions be reconstructed? | Can material termination rationales, multiple causes, uncertainty, rule versions, overrides and later changes be reviewed by an authorized party? |
| 5. Evidentiary integration | How do in vitro, in silico, mechanistic, animal, clinical and real-world evidence converge? | What sufficiently independent evidence addresses the evidence-support trigger, and what evidence demonstrates functional replacement without treating the historical rule as ground truth? |
| 6. Learning and revision | How is performance monitored and the use claim updated? | Do changes in generator priors, objectives, filters, thresholds, corroboration, reachable chemistry or model authority trigger reassessment? |

### 6.3 Generate targeted technical and challenge evidence

Challenge evidence addresses the newly reachable chemistry and the displaced biological function if a credible function was found. Where feasible, we include informative terminated compounds, matched pairs across historical guardrail boundaries, prospectively generated candidates, non-safety provenance controls and negative controls. We blind the evaluations and seek orthogonal evidence with failure modes distinct from those of the model and generator. Historical archives may provide informative evidence (Raccuglia et al., 2016; Sanz et al., 2017; Shinozawa et al., 2025), but where they do not, prospective testing and matched comparisons become critical. The technical characterization traces a candidate from proposal through synthesis, screening and advancement, and then through definitive evaluation, termination or retention, and reporting. On its own, an inventory of the final training set cannot reveal all upstream selection effects; thus, we record the reasons for an exclusion together with the supporting evidence, mechanisms and thresholds, and the rule versions, confidence, competing causes and any subsequent overrides. Standard model characterization remains necessary as well because a record is only useful if someone can check it.

### 6.4 Establish auditable provenance

Data provenance documents how observations were generated and handled. Exclusion provenance explains why the observations were never generated or why their results were unavailable. A provenance register extends established documentation practices to these material absences, and to the selection decisions behind them (OECD, 2025; Gebru et al., 2021; Mitchell et al., 2019). Importantly, REGARD does not require that proprietary structures or attrition rationales be publicly disclosed. A confidential review, independent audits, secure data environments, aggregate reporting and blinded challenge studies may each provide proper access while the commercial information is protected. The relevant knowledge is distributed among multiple parties. Cross-organizational initiatives have demonstrated that selected forms of data aggregation are feasible (Waring et al., 2015; Sanz et al., 2017). At a minimum, a claim of consequential use must be auditable by an authorized party who has access to the supporting evidence because the record exists to be evaluated, not to decide.

### 6.5 Integrate evidence to justify model reliance

Historical decisions identify the regions and the mechanisms that require evaluation. They do not determine the outcome of that evaluation. The evidence may include assays on human cells, microphysiological systems, models of the exposure and mechanistic biomarkers, and it may include causal models, clinical or pharmacovigilance data and informative animal findings (ICCVAM, 2024; FDA, 2026b). We document the conclusions on the evidence support and on the functional replacement separately and judge a replacement by the coverage of the relevant biological risk, and not by its agreement with the historical exclusions. For example, a multichannel cardiac assessment may improve discrimination beyond screening based on hERG alone (Kramer et al., 2013), and integrated liver models may give a better account of metabolism, mitochondrial dysfunction and bile acid transport, as well as exposure and immune mechanisms. Yet neither human biological relevance nor assay diversity establishes sufficient coverage or independence. Unresolved gaps and conflicting findings limit the authority assigned to the model's predictions, and such judgments hold only while the system stays as it is.

### 6.6 Establish lifecycle monitoring and reassessment

Reachable chemistry can be altered by changes in generative priors, optimization objectives, constraints and filters (Fannjiang et al., 2022; Langevin et al., 2023; Yoshizawa et al., 2025). We also propose monitoring changes in building blocks, query budgets and human overrides. We prespecify which changes require a reassessment, which include expanded reachability of a guardrail domain, altered corroboration and increased model authority. Importantly, in REGARD they all apply even when the safety model remains unchanged, extending the lifecycle monitoring and revalidation triggers proposed for NAMs (Kopańska and Hartung, 2026). Our monitoring distinguishes a change in predictive performance from a change in the evidence system. We track crossings of historical guardrails, independent safety outcomes and discordant cases. Thus, a reassessment may justify expanding, restricting, suspending or withdrawing the authorized use, and if so, we record each of these decisions and the authority responsible for it.

Thus, REGARD should facilitate scientific progress without preserving obsolete safety rules. When a workflow re-enters a historically restricted domain, documentation should specify the rationale for the change, the credibility of the historical safety function, evidence supporting its retention or replacement, the responsible decision authority, and the monitoring plan. Improved human-relevant methods may supersede historical safeguards when they provide equivalent or superior assessment of the relevant biological risks.

## 7 Limitations

Historical rules may be incorrect, overinclusive or obsolete. Even rigorous rules may have incomplete provenance because termination decisions are proprietary, inconsistently recorded or unavailable. Published data describe reasons for terminating nominated candidates (Waring et al., 2015) and controlled access to legacy safety reports (Sanz et al., 2017), but do not quantify undocumented proposal-stage exclusions. Nor does a complete record settle attribution when attrition has multiple causes. Even where attribution succeeds, definitive human outcomes may be ethically unobtainable, requiring orthogonal human-relevant and mechanistic evidence while leaving residual uncertainty.

REGARD addresses a specific problem of evidence transfer, not every form of distribution shift. We must evaluate the contribution of exclusion provenance against a complete baseline of conventional controls. Where provenance-specific requirements are unsupported or add no decision-relevant benefit, they should be narrowed or rejected. These limitations are not a reason to wait. The triggers can serve as screening conditions, so where both are known to be zero and conventional controls support the intended use, REGARD requires nothing beyond those controls and the initial provenance audit. A trigger value of zero does not establish that a candidate is safe. When a trigger is met, the adequacy of conventional controls must be reassessed, since reweighting alone cannot recover outcomes that were never observed. Addressing the two triggers while workflows are being designed allows developers and safety assessors to determine if conventional controls suffice, additional evidence is needed, or a safety function must be preserved.

## 8 REGARD empirical evaluation

To determine if historical exclusion provenance improves the identification and management of safety risks, we propose to evaluate REGARD in DILI, followed by replication in cardiac liability (Fig. 3). The historical guardrail, reconstructed domain, assays, materiality thresholds, minimum meaningful improvement, primary decision outcome and analysis plan are preregistered, and the safety NAM and the decision policies are locked before any evaluation outcome is examined. Both policies share the candidate distribution, the fixed NAM, the permitted actions and the evidence budget, so, apart from any cost differences explicitly included in the decision loss, they differ only in the use of exclusion provenance.

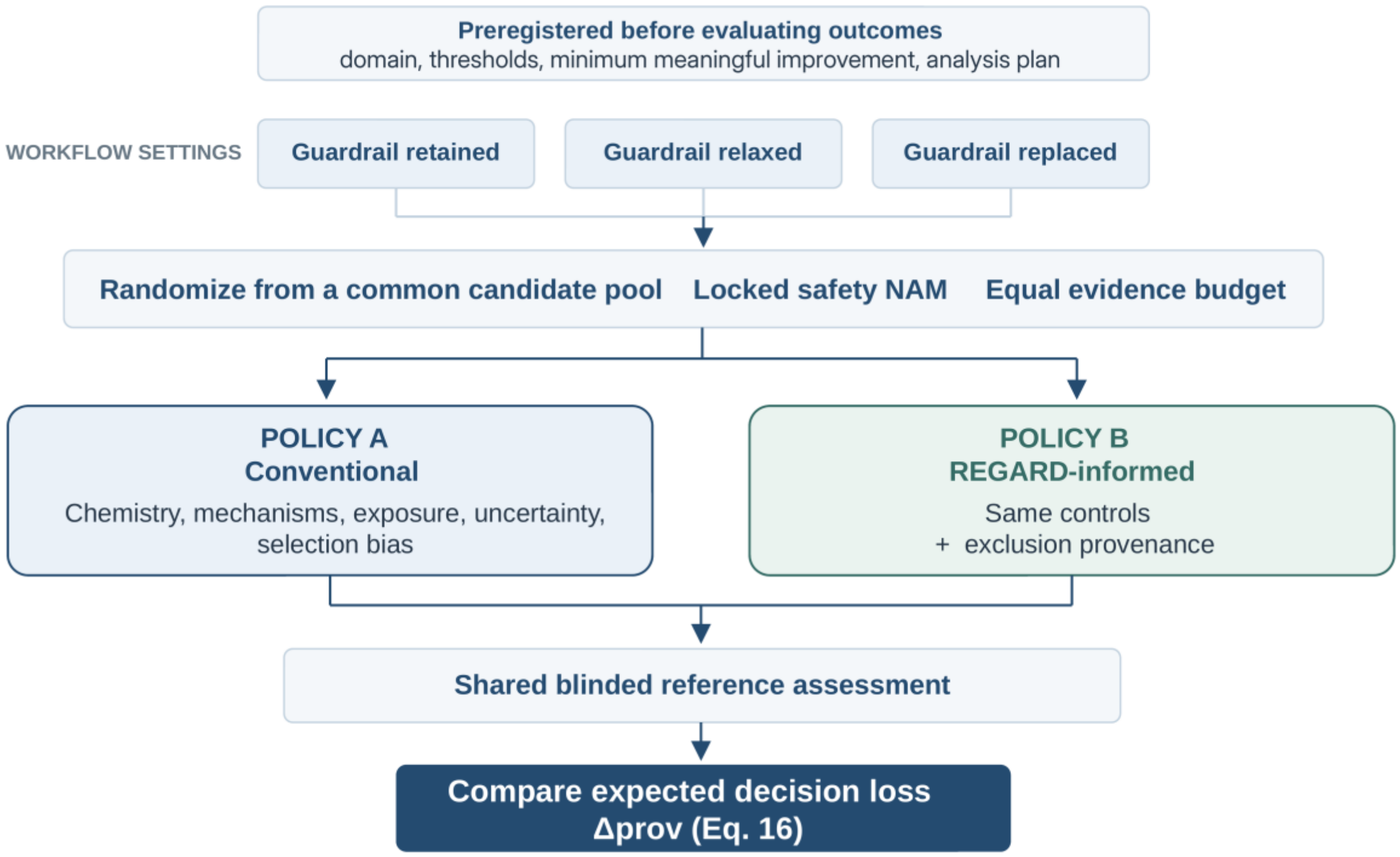

**Fig. 3: Proposed evaluation of REGARD.** Candidates are sampled from workflows that retain, relax, or independently replace a historical guardrail. Both policies then share the candidate distribution, the fixed NAM, the permitted actions and the evidence budget, so that, apart from any cost differences included explicitly in the decision loss, they differ only in the use of exclusion provenance. $\Delta_{\text{prov}}$ is the incremental decision value of provenance defined in Equation 16. The shared blinded reference assessment may use a prespecified verification sample with known, positive inclusion probabilities (Supplementary Methods S6).

Formally, let $\pi$ denote a fully specified decision policy, including provenance inputs, assay requests, referral, abstention, and any randomization. Let $\ell_{\text{dec}}^{\pi}(x, y)$ be its conditional expected total decision loss given candidate characteristics $X = x$ and reference outcome $Y = y$, averaging over baseline provenance and the measurement and action histories induced by $\pi$. Policies share the joint distribution of $X$, $Y$, and baseline provenance and the conditional measurement-response mechanisms but may induce different histories. The loss includes consequences and costs of all permitted actions, including referral and abstention. Assume finite expected loss and a common reference outcome unaffected by policy assignment.

Under a common candidate distribution $P_G$ and endpoint context $C_G$, define the expected policy loss:

$$\mathcal{R}_G(\pi) = \mathop{\mathbb{E}}_{\substack{X \sim P_G \\ Y \sim P^*(\cdot|X, C_G)}} [\ell_{\text{dec}}^{\pi}(X, Y)] \qquad \text{(Eq. 15)}$$

The primary estimand is the incremental value of adding provenance to the specified decision policy:

$$\Delta_{\text{prov}} = \mathcal{R}_G(\pi_{\text{conv}}) - \mathcal{R}_G(\pi_{\text{conv+prov}}) \quad \text{(Eq. 16)}$$

Animal use is a measured secondary outcome, so the contribution of a provenance-informed policy to the 3Rs is evaluated rather than assumed. A positive and reproducible contrast in Equation 16 indicates a reduced decision loss for the specified population, endpoint and loss function. A precisely estimated null, or an interval that excludes a meaningful benefit, would support narrowing or rejecting the provenance-specific requirements in the context tested. An imprecise null does not establish that no benefit exists. For sampling design, identification conditions, off-policy estimators (Dawid and Constantinou, 2014; Jiang and Li, 2016), transport weights, proxy-outcome contrast and external-validation requirements see Supplementary Methods S5 and S6.

## 9 Conclusions

Historical safety decisions shape both the evidence available to models and the safeguards surrounding their use. As generative AI brings previously excluded chemistry back into discovery, the evidence supporting a model may no longer justify the decisions it is asked to make and safeguards that once supported those decisions may no longer be in place. This can happen even when the model itself remains unchanged.

To determine whether continued reliance on a model is justified, REGARD addresses two distinct problems, the gaps in outcome evidence and the loss of biological safety functions. The distinction matters because a replacement assay cannot recover missing historical outcomes, and additional outcome data do not establish that a safety function remains covered. The two problems require different evidence and different remedies.

Distinguishing between these problems also creates opportunities for scientific progress. REGARD helps determine which historical rules are still needed. Historical rules that lack scientific support can be retired, while those serving a credible safety function can be replaced by more effective human-relevant methods. Where the evidence supports such a replacement, animal studies can be avoided; where no adequate alternative exists, their continued use can be justified. REGARD therefore provides a basis for reducing animal use without compromising biological safety assessment or preserving outdated practices.

As new discovery workflows are being designed, we have an opportunity to reconsider not only how safety is assessed but also what evidence justifies our decisions. A model may inherit its credibility from the past, but its authority must be justified in the workflow of the future.

## Declarations

**Conflict of interest:** The authors declare no conflicts of interest.

**Data availability:** No new datasets were generated or analyzed for this Concept Article.

**Acknowledgements:** This work was supported by the Chair's Mid-Career Genetics Faculty Research Fund Award, Department of Genetics, University of North Carolina at Chapel Hill.

**Use of generative AI:** The authors used Claude (Opus 5.5, Anthropic) to assist with consistency checking, mathematical equation rendering, and manuscript formatting.

**Author contributions:** Andrew Dunn: Formal analysis, writing, review and editing. Adriana S. Beltran: Conceptualization, writing, review and editing.

## Supplementary Methods S1. Stage-resolved selection model, positivity, and identification conditions

As described in Section 2.2 of the main text. The notation and Equations 1 to 4 are given in the main text; this supplement states the selection mechanism they are defined over, the conditions under which the components of Equation 4 are identifiable, and the sense in which the attribution between its first two terms is arbitrary.

### S1.1 The selection mechanism

The historical joint law $Q_H$ is determined by the proposal distribution $P_H$, the outcome law $P^*(Y \mid X, C_H)$, and a sequence of stage-specific selection and reporting mechanisms. Write the probability that a candidate passes stage $k$, given its features and its history, as

$$\tau_k(x, h_{k-1}) = Q_H(S_k = 1 \mid X = x,\, S_1 = \cdots = S_{k-1} = 1,\, H_{k-1} = h_{k-1}) \qquad \text{(Eq. S1-1)}$$

where $k = 1, \dots, K$ indexes the stages, $S_K$ is the final stage of definitive evaluation, and $H_{k-1}$ collects information available before stage $k$, including early assay results, structural alerts, reactive-metabolite findings and expert judgments ($H_0$ is empty). Write $S_0 = 1$ for entry into the proposal population. For candidates with $X = x$, define the survivor-conditional passage probability $a_k(x) = Q_H(S_k = 1 \mid X = x, S_{k-1} = 1)$. It is the average of $\tau_k(x, H_{k-1})$ over the history distribution among candidates who reached stage $k$. Let $r(x) = Q_H(O = 1 \mid X = x, S_K = 1)$, which averages reporting over the histories and outcomes of definitively evaluated candidates. By the chain rule and the nesting of the stages,

$$Q_H(O = 1 \mid X = x) = r(x) \prod_{k=1}^{K} a_k(x) \qquad \text{(Eq. S1-2)}$$

Equation S1-2 uses survivor-conditional averages at each stage, not probabilities conditional on the entire realized history. If any passage probability is zero, the observation probability is zero; later conditional factors need not be defined on that null event. An early stage with a small passage probability can dominate the availability of evidence, however permissive later stages are. Moreover, $\tau_k$ may depend on outcome-relevant components of $H_{k-1}$ that were never recorded, as may reporting. Such mechanisms can prevent recovery of outcome evidence from the selected archive alone.

### S1.2 Shift types and the role of positivity

The divergence between the observed and the intended-use risk arises from two shift types and one observation mechanism. Under covariate shift, $P(X)$ changes while $P(Y \mid X)$ is invariant. Under concept or mechanism shift, $P^*(Y \mid X, C)$ differs between the historical and the target context even at fixed $X$, because the exposure, the formulation, the population, the assay or the biology has changed (Quiñonero-Candela et al., 2008). Separately, under selective labels or missing-not-at-random observation, the availability of a label depends on prior decisions (Lakkaraju et al., 2017).

Positivity is not a third shift type. It determines whether the selection mechanism can be corrected for the population of interest. For identification of historical proposal-stage risk, the requirement is

$$Q_H(O = 1 \mid X = x) > 0 \quad \text{for } P_H\text{-almost every } x \qquad \text{(Eq. S1-3)}$$

For transfer of historical outcome evidence to the intended-use population, one additionally needs $P_G \ll P_H$ (the intended distribution is absolutely continuous with respect to the historical proposal distribution) and positive historical observation probability for $P_G$-almost every $x$, together with appropriate outcome-context transport assumptions. Zero observation probability on a region of positive mass in the relevant population precludes nonparametric recovery from the observed evidence alone; near-zero probability can instead make estimation unstable (Petersen et al., 2012).

### S1.3 When reweighting identifies the historical selection term

The third bracket of Equation 4, $R_H(m) - R_{\text{obs}}(m)$, is identifiable by inverse-probability weighting under the following conditions, which correspond to the standard missing-at-random argument specialized to this setting (Liu and Kashima, 2022).

(i) Observed selection: The historical marginal observation probability $Q_H(O = 1 \mid X)$ is known or consistently estimable from suitable proposal-stage records or other evidence. Stage-specific information may support this estimation but is not by itself sufficient.

(ii) *Conditional ignorability:* $O$ is independent of $Y$ given $X$, that is $Q_H(O = 1 \mid X, Y) = Q_H(O = 1 \mid X)$. Equivalently, once $X$ is conditioned on, selection carries no further information about the outcome.

(iii) Positivity: Equation S1-3 holds on the historical proposal population. For stable estimation, observation probabilities should additionally be bounded away from zero or the resulting weight instability should be explicitly assessed.

Under (i) to (iii),

$$R_H(m) = \mathbb{E}_{Q_H}\left[\frac{L(m(X),Y)\,\mathbf{1}\{O=1\}}{Q_H(O=1|X)}\right] \qquad \text{(Eq. S1-4)}$$

and the selection term is recovered.

Each condition may fail in safety-motivated exclusion. Condition (i) may fail when the records needed to estimate observation probabilities are missing. Condition (ii) fails when exclusion depends on outcome-relevant information not adequately represented in $X$, such as unrecorded mechanistic findings or expert judgments. Where a structural alert is a deterministic function of $X$, conditional ignorability need not fail, although consistent exclusion based on that alert may violate positivity (Lakkaraju et al., 2017). Condition (iii) fails where the guardrail prevents observation through every available route. At zero observation probability, recovery requires assumptions or evidence external to the selected archive (Heckman, 1979; Bareinboim et al., 2014). Near-zero probabilities produce unstable weights (Cole and Hernán, 2008; Petersen et al., 2012), and partial identification may be preferable to an unsupported point estimate (Manski, 2003).

### S1.4 Order-dependence of the decomposition

Equation 4 introduces the cross-setting risk $R_{G,H}(m)$, which applies the intended proposal distribution under the historical outcome context. The mirror-image construction applies the historical proposal distribution under the target context,

$$R_{H,G}(m) = \mathop{\mathbb{E}}_{\substack{X \sim P_H \\ Y \sim P^*(\cdot|X, C_G)}} [L(m(X), Y)] \quad \text{(Eq. S1-5)}$$

and yields the equally valid identity

$$R_G(m) - R_{\text{obs}}(m) = \left[R_G(m) - R_{H,G}(m)\right] + \left[R_{H,G}(m) - R_H(m)\right] + \left[R_H(m) - R_{\text{obs}}(m)\right] \quad \text{(Eq. S1-6)}$$

The first bracket of Equation S1-6 is a generator shift evaluated under the target context and the second is a context shift evaluated under the historical proposal distribution, so the allocation between the two differs from that of Equation 4 in the main text. The total contrast and the third bracket are identical in both. Two consequences follow. The attribution of a given amount of risk change to “context” rather than “generator” is not a property of the system but of the order in which the two are introduced, and the two decompositions agree exactly when the shifts do not interact, that is when $R_G(m) - R_{G,H}(m) = R_{H,G}(m) - R_H(m)$. Reported decompositions should therefore state which ordering was used, and the selection term, which is the term REGARD acts on, is the one that is invariant to the choice.

### References cited in this supplement

## Supplementary Methods S2. Reachability and enrichment: estimators, confidence procedures, and correction for domain misclassification

Equations 5 to 11 are defined in the main text. This supplement describes the sampling model underlying the estimators in Equations 8 and 11, the one-sided testing used for the rule-out test, and the correction required when membership in $A_g$ is assigned by a reconstructed classifier rather than observed.

### S2.1 Sampling model

Let $n_G$ candidates be drawn from the intended generator and $n_{H,B}$ records be obtained from the historical archive, and let the corresponding counts within the guardrail domain be $y_G$ and $y_{H,B}$. Treating the two samples as independent binomials, $\hat{p}_G = y_G/n_G$ and $\hat{p}_{H,B} = y_{H,B}/n_{H,B}$, which gives the estimators of Equations 8 and 11. The archive sample is not a random sample of historical proposals; therefore, Equation 7 is a bound and not an estimate. The binomial model captures sampling variability within the archive, but does not account for the selection mechanism through which the archive was generated.

### S2.2 One-sided interval for the absolute bound

The rule-out test described in the main text requires a one-sided upper confidence limit for $\overline{\Delta}_g = p_G - p_{H,B}$. For the difference between two independent proportions, we recommend the Newcombe hybrid score interval, which generally provides better coverage than the Wald interval, particularly at low prevalences and near the boundaries of the parameter space (Newcombe, 1998). Denoting its upper limit by $U_{1-\alpha}$, a prespecified material increase $\delta^*$ is ruled out when

$$U_{1-\alpha}\left(\overline{\Delta}_g\right) < \delta^* \quad \text{(Eq. S2-1)}$$

provided that the monotone-depletion assumption in Equation 6 is defensible. The test is one-sided by construction. Failure to rule out a material increase does not imply that such an increase occurred, nor does a large value of $U_{1-\alpha}$ establish displacement.

### S2.3 One-sided interval for the enrichment bound

For the ratio $\overline{\rho}_g = p_G/p_{H,B}$ we recommend a score-based interval (Koopman, 1984) or a profile-likelihood interval instead of the delta-method interval, which is unreliable when $p_{H,B}$ is small. When the observed archive count $y_{H,B} = 0$, the empirical ratio is undefined or non-finite; a zero count does not establish zero population prevalence. The observed data alone do not support a finite upper confidence limit in this case. We report the counts and a one-sided interval for the absolute difference, without reporting a ratio. We do not use continuity corrections or added pseudocounts to manufacture a finite estimate.

### S2.4 Correction for non-differential domain misclassification

Let $A$ denote true membership in $A_g$ and $\hat{A}$ the assignment made by the reconstructed classifier of Section 3.1, with sensitivity $se = P\left(\hat{A} = 1 \mid A = 1\right)$ and specificity $sp = P\left(\hat{A} = 0 \mid A = 0\right)$.

Assume non-differential misclassification, that is, $se$ and $sp$ are the same in the generator sample and the archive and do not depend on $Y$. For any population with true prevalence $p$, the apparent prevalence is

$$p^* = P(\hat{A} = 1) = se\, p + (1 - sp)(1 - p) = p(se + sp - 1) + (1 - sp) \quad \text{(Eq. S2-2)}$$

Applying Equation S2-2 to both samples, the additive error term cancels in the difference and

$$p_G^* - p_{H,B}^* = (se + sp - 1)(p_G - p_{H,B}) = (se + sp - 1)\overline{\Delta}_g \quad \text{(Eq. S2-3)}$$

When $0 < se + sp - 1 < 1$, the apparent absolute bound is attenuated toward zero. For a positive absolute bound, the apparent contrast is smaller than the true upper bound. Consequently, when testing against a positive materiality threshold, ignoring misclassification can render the rule-out procedure in Equation S2-1 anti-conservative. For a negative bound, attenuation moves the contrast upward toward zero. This distinction is reflected in the main text.

For the ratio, substituting Equation S2-2 into the population ratio in Equation 10 gives

$$\rho_g^* = \frac{p_G(se+sp-1)+(1-sp)}{p_{H,B}(se+sp-1)+(1-sp)} \quad \text{(Eq. S2-4)}$$

which tends to one as (1 − sp) grows relative to the true prevalences. Imperfect specificity therefore dilutes apparent enrichment, and does so most severely in exactly the regime of interest, where $p_{H,B}$ is small.

Prevalences are corrected by the Rogan-Gladen estimator (Rogan and Gladen, 1978)

$$\hat{p} = \frac{\hat{p}^*-(1-\widehat{sp})}{\widehat{se}+\widehat{sp}-1} \quad \text{(Eq. S2-5)}$$

truncated at zero and one, with se and sp estimated from a validation subset of candidates whose true domain membership is established by expert review. Uncertainty in se and sp must be propagated rather than treated as fixed: Equation S2-5 is a ratio of estimated quantities and its variance is inflated substantially when se + sp − 1 is near zero. We recommend a nonparametric bootstrap that resamples the generator sample, the archive sample and the validation subset jointly, and we note that the corrected interval can be wide enough to be uninformative when the classifier is weak. That outcome should be reported as an uninformative reconstruction rather than as an absence of re-entry. If validation uncertainty includes classifiers with nonpositive discrimination, the corrected contrast is not reliably identified by this inversion; report uncertainty or use a justified partial-identification procedure rather than discarding such replicates.

**References cited in this supplement**

## Supplementary Methods S3. Reconstruction of guardrail domains and propagation of classification error

This supplement describes how a historical guardrail domain $A_g$ is reconstructed when the rule was not structural, how the reconstruction is characterized, and how its error is carried into the quantities of Section 3.2 and the decision triggers of Sections 3.3 and 3.4.

### S3.1 Three kinds of domain

A structural rule defines $A_g$ directly and does not require reconstruction. Membership is a deterministic function of the molecular graph, and the only documentation required is the rule text, its version and its active period. An exposure-dependent rule defines $A_g$ only in conjunction with assumed exposure conditions, and membership must therefore be evaluated under prespecified conditions which are reported with the domain. A judgment-based practice does not explicitly define a domain and must be reconstructed as a classifier.

### S3.2 Reconstructing a judgment-based domain

We reconstruct a judgment-based practice as a classifier Â(x) trained or elicited on candidates whose historical selection decisions are documented. We report all estimates, together with the time period and sponsor to which the reconstruction applies, the rule version and any superseding versions. We record the inputs available to the original decision-makers, which define the information that can be used for a faithful reconstruction, together with the thresholds applied and any documented overrides. We also report the inter-rater agreement among the original decision-makers, which informs the interpretation of reconstruction agreement, and the sensitivity se and specificity sp of the reconstruction evaluated using a held-out set of documented decisions.

Two constraints on the reconstruction are worth describing. It must not use information that postdates the decision, including the outcome being predicted, since a classifier that has seen the outcome will reconstruct a better rule than the one that was actually applied. Where inter-rater agreement among the original decision-makers was low, a high-accuracy reconstruction is potential evidence of leakage rather than fidelity.

### S3.3 Probabilistic membership

Where the classifier returns a calibrated probability rather than a label, domain prevalence is the expectation of the population-specific membership probability. Let $P_j$ denote the proposal distribution for $j \in \{H, G\}$, and use the conditional archive distribution for the archive prevalence:

$$p_j = \mathbb{E}_{X \sim P_j}[s_j(X)], \quad s_j(x) = P_j(A = 1 \mid X = x) \quad \text{(Eq. S3-1)}$$

For the archive, $p_{H,B} = \mathbb{E}_{Q_H}[s_{H,B}(X) \mid B = 1]$, where $s_{H,B}(x) = Q_H(A = 1 \mid X = x, B = 1)$. Estimated scores must be calibrated in each relevant population or justified as transportable; a historical held-out set alone does not establish calibration in newly reachable chemistry. Probabilistic membership is preferred to thresholding where the score is calibrated, because thresholding discards information near the domain boundary. That distinction is material to the referral and abstention decisions of Section 6.5. The binary misclassification correction in S2.4 applies to hard classifications, not automatically to calibrated probability scores.

### S3.4 Overlapping domains and competing reasons

A candidate may belong to several guardrail domains, and a historical exclusion may have had several causes. Neither is reduced to a single reason. Where domains overlap, the reachability quantities of Section 3.2 are computed per domain and the triggers of Sections 3.3 and 3.4 are evaluated per domain, since a credible safety function may attach to one of two overlapping rules and not the other. Where an exclusion had several recorded causes, the candidate contributes to the safety-related stratum and to each competing stratum, and we report the sensitivity of the conclusion to that allocation.

### S3.5 Propagation into the triggers

Classification error affects the triggers through $E_g$, which depends on reachability and decision exposure as defined in Section 3.2. Apply the correction in S2.4 to hard-label marginal prevalence estimates, and the joint-probability correction in S4.3 to decision exposure. For calibrated probability scores, propagate calibration and transport uncertainty instead. Prespecify the materiality thresholds and one-sided confidence levels before analysis (for example, 95% one-sided limits with adjustment for multiple decision criteria where needed). For an identified contrast, materiality is established when its lower confidence limit exceeds the threshold and ruled out when its upper confidence limit lies below it; otherwise it is unresolved. An archive-based upper bound can rule out material proposal-stage re-entry under monotone depletion but cannot establish it, even if the bound is large. Evaluate the prespecified reachability and decision-exposure criteria jointly: $E_g = 1$ if at least one is established, $E_g = 0$ if all are ruled out, and otherwise $E_g$ is unknown. Apply the conjunction logic of Section 3.3: an unresolved $E_g$ makes the trigger unknown only if no other required condition is known to fail. A reconstruction too weak to resolve materiality is not evidence of absence of re-entry.

## Supplementary Methods S4. Consequential decision exposure: identification requirements and the per-entrant contrast

As described in Section 3.2 of the main text, where Equation 12 defines the expected decision counts $\mu_{g,j}$ and their contrast $\Delta_g^{dec}$. This supplement describes the conditions required to identify Equation 12, and defines the per-entrant counterpart used when throughput differs between the historical and the intended workflow.

### S4.1 What Equation 12 requires

Equation 12 combines a candidate count with a joint probability. Identification of the expected decision counts requires the following conditions.

**(i) A defined accounting period or budget:** $N_j$ is the number of candidates entering workflow j over a prespecified period or budget. The accounting period must be fixed in advance and common to both processes, since otherwise $\Delta_g^{dec}$ compares different accounting windows.

**(ii) An observable decision point:** $V_j$ = 1 must correspond to a recorded decision governed by the model, as defined in Section 5, and representing the same decision in both processes. Where the intended workflow introduces a decision point that has no historical counterpart, $\mu_{g,H}$ is zero by construction. The contrast then measures the creation of exposure rather than an increase at an existing decision point. This distinction must be reported explicitly, rather than interpreted as a large increase in existing exposure.

**(iii) Joint rather than marginal measurement:** $Q_j(X \in A_g, V_j = 1)$ is a joint probability and cannot generally be estimated by multiplying the proposal-stage prevalence by the overall decision rate. Candidates within a guardrail domain may be filtered at different rates than those outside it. Consequently, a proposal prevalence of 15% alone does not determine the number of affected decisions.

For the historical process, conditions (i) to (iii) can be satisfied only where workflow records survive at the decision stage. Where they do not, $\mu_{g,H}$ is not identifiable. Thus, the comparison should be reported as a statement about the intended workflow alone, with the historical term treated as unavailable instead of zero.

### S4.2 The per-entrant contrast

The counts in Equation 12 scale with throughput, meaning that a workflow processing more candidates will produce a larger $\mu_{g,G}$ even when selectivity remains unchanged. Where throughput differs, or where it is expected to change over the period of interest, the per-entrant quantity separates selectivity from volume:

$$q_{g,j} = \frac{\mu_{g,j}}{N_j} = Q_j\left(X \in A_g, V_j = 1\right), \quad \Delta_g^{\mathrm{per}} = q_{g,G} - q_{g,H} \quad \text{(Eq. S4-1)}$$

We report both quantities. The per-entrant contrast in Equation S4-1 indicates whether the intended workflow is more likely to bring a guardrail-domain candidate to a consequential decision, whereas the absolute contrast in Equation 12 indicates the expected difference in the number of such decisions. Materiality depends on both, and they can diverge. A workflow may be less selective per entrant while still generating far more affected decisions, because its throughput is higher. Conversely, a workflow with a large per-entrant increase may generate few affected decisions in absolute terms. Neither quantity alone establishes a trigger, for the reasons given at the end of Section 3.2.

### S4.3 Uncertainty

For an administrative budget, $N_j$ is fixed and Equation 12 applies. If throughput is random, let $q_{g,j}(n)$ be the conditional expected fraction of entrants within the domain reaching the consequential decision given $N_j = n$ (set the count contribution to zero when $n = 0$). The expected count is then $\mu_{g,j} = \mathbb{E}[N_j q_{g,j}(N_j)]$. In general this is not the product of expected throughput and a run-averaged decision probability. For random throughput, an entrant-weighted exposure probability is $\mu_{g,j}/\mathbb{E}[N_j]$ when the denominator is positive; distinguish this from the mean of run-specific proportions. We propagate joint uncertainty in throughput and candidate composition and use cluster resampling at the level of the generator run or archive sponsor where appropriate. For reconstructed hard-label domain membership, let $v_j = Q_j(V_j = 1)$ and $q^*_{g,j} = Q_j(\hat{A} = 1, V_j = 1)$. If sensitivity and specificity are constant across workflows conditional on reaching the decision point, then

$$q^*_{g,j} = (se + sp - 1)q_{g,j} + (1 - sp)v_j \quad \text{(Eq. S4-2)}$$

$$q_{g,j} = \frac{q^*_{g,j} - (1 - sp)v_j}{se + sp - 1} \quad \text{(Eq. S4-3)}$$

provided $se + sp - 1 > 0$. Estimate classification accuracy among candidates reaching the decision and propagate uncertainty in $se$, $sp$, $v_j$ and the observed joint probability. The apparent per-entrant contrast contains an additional term $(1 - sp)(v_G - v_H)$, so attenuation toward zero is not guaranteed. When throughput is random, apply the correction conditionally or to the appropriately entrant-weighted population. If the conditional classification accuracy cannot be justified, report sensitivity analyses or bounds rather than asserting attenuation.

## Supplementary Methods S5. Expected policy loss: definitions, identification, and transport to the intended-use population

As described in Section 8 of the main text, where Equation 15 defines the expected policy loss $\mathcal{R}_G(\pi)$ and Equation 16 the incremental value of provenance $\Delta_{prov}$. Here we describe the policy and loss definitions, the conditions under which Equation 16 is identifiable, and the weighting required when the evaluated candidates do not represent $P_G$.

### S5.1 Policies and conditional loss

The policy and loss definitions are as given in Section 8 of the main text. For a deterministic direct-action policy using only $X$, $\ell^{\pi}_{\text{dec}}(x, y) = L_{\text{dec}}(\pi(x), y)$.

### S5.2 Identification of Equation 15

Expected policy loss is identified in either of two ways.

Randomized assignment. Policies are assigned at random to candidates or to batches, with the assignment implemented as specified and outcomes ascertained for both arms. This identifies $\mathcal{R}_G(\pi)$ directly and is the preferred design, because it requires no assumption about the measurement histories the policies induce.

Replay. We evaluate a single set of candidates under a common reference assessment and apply each policy retrospectively to the recorded measurements, actions and costs. Replay identifies $\mathcal{R}_G(\pi)$ only under a response-invariance assumption, namely that the measurements a policy would have requested are recoverable from those actually taken and that the response to a measurement does not depend on the policy which requested it. A common reference endpoint alone does not identify the unobserved policy-specific histories. Two policies that request different assays reach the reference outcome through different paths, and only the endpoint is shared.

Where neither design is available, observational off-policy evaluation requires a prespecified identification strategy, for example weighting under consistency, sequential exchangeability and positivity, with assignment probabilities either known or identified (Jiang and Li, 2016). We specify the strategy before examining the data, and we report the estimator, the weights and the overlap diagnostics.

### S5.3 Transport to the intended-use population

Candidates assembled for a challenge study are typically matched on domain distance, predicted risk, uncertainty and synthesis feasibility, and enriched for guardrail-domain membership. Such a set need not represent the intended proposal distribution $P_G$. Let $\ell^{\pi}_{\text{dec,chal}}(x, y)$ denote the conditional expected loss in the challenge population, and let $Q_{\text{chal}}$ denote its joint distribution of $X, Y$ in context $C_G$. The unweighted challenge-population contrast is

$$\Delta^{\text{chal}}_{\text{prov}} = \mathbb{E}_{(X,Y)\sim Q_{\text{chal}}}\left[\ell^{\pi_{\text{conv}}}_{\text{dec,chal}}(X, Y) - \ell^{\pi_{\text{conv+prov}}}_{\text{dec,chal}}(X, Y)\right] \quad \text{(Eq. S5-1)}$$

which is a different estimand from Equation 16. Estimation for $P_G$ requires either representative sampling from the generator or transport weights

$$w(x) = \frac{dP_G(x)}{dP_{\text{chal}}(x)} \quad \text{(Eq. S5-2)}$$

with adequate overlap (the intended distribution must be absolutely continuous with respect to the challenge distribution) and sufficiently stable weights. Weighting on $X$ identifies the intended-use contrast only if the conditional expected policy-loss difference given $X$ is transportable between populations. This requires appropriate outcome-context and policy-response assumptions, including the distribution of baseline provenance conditional on $X$. If transportability requires additional baseline variables, such as exposure or provenance, use their joint density ratio instead. Where overlap, weight stability or conditional transportability is inadequate, we report Equation S5-1 separately and do not describe it as the intended-use contrast. Reporting the challenge-population result as though it were Equation 16 may misrepresent the intended-use value of provenance. The direction and magnitude of this discrepancy depend on the enrichment strategy and the heterogeneity of the policy-loss contrast across candidate strata.

### S5.4 Leakage, dependence, and multiplicity

The policy inputs for each candidate must be available before its corresponding held-out endpoint is assessed. Provenance must not contain the endpoint being predicted, nor a termination rationale derived from the same outcome assay, since either would create label leakage and could produce a spurious apparent benefit. Historical outcomes from separate development compounds may inform the reconstruction of provenance, provided that the separation of the data is maintained and auditable. Candidates may be nested within guardrails, generators, predictors, chemotypes and sponsors, so the analysis uses a hierarchical model, a cluster bootstrap or an equivalent method that respects this dependence. One primary estimand is prespecified, and secondary analyses are subject to multiplicity control.

### S5.5 Proxy outcomes

When the definitive outcome Y cannot be observed prospectively, a human-relevant proxy W is preregistered along with a separate loss contrast defined using that proxy. Importantly, an improvement for the proxy does not necessarily imply an improvement for the definitive outcome. Inference from one to the other requires a justified measurement or transport model, explicit bounds, or a sensitivity analysis. Neither biological plausibility nor preregistration is sufficient to establish this relationship.

The decision loss gives explicit weight to severe false-negative consequences, and calibration, abstention efficiency, assay-selection utility, false-positive exclusion, delay, cost and animal use serve either as secondary outcomes or as components of the loss. Both policies use the same locked NAM, the same permitted action set and the same evidence budget, unless the differing costs are themselves included in the loss.

### References cited in this supplement

## Supplementary Methods S6. Evaluation protocol for the direct empirical test

With the design fixed, we draw candidates from workflows that retain, relax or independently replace the historical guardrail, using several generators and predictors. To account for different candidate populations, sampling covers prespecified strata of chemistry, mechanism, exposure, potency, uncertainty and synthesis feasibility, together with non-safety provenance controls and negative controls. Where feasible, we include candidates rejected or referred by either policy in a blinded common reference assessment. If outcome ascertainment is impractical, we use a prespecified verification sample with known inclusion probabilities positive in every relevant stratum and justified weighting or transport assumptions, to reduce verification bias while controlling experimental costs.

Our comparison isolates the incremental decision value of exclusion provenance against conventional controls for chemistry, mechanism, exposure, uncertainty and selection bias. Provenance may inform stratification, referral, abstention, limits on model authority or the selection of corroborating evidence. Both policies use the same locked NAM, permitted actions and evidence budget, except where cost differences are explicitly included in the decision loss.

The conditional decision-loss definitions, including the deterministic direct-action special case, are given in Supplementary Methods S5.1.

The expected policy loss in Equation 15 can be identified by randomized assignment or justified replay, as detailed in Supplementary Methods S5.2. Observational off-policy evaluation requires a prespecified identification strategy (Dawid and Constantinou, 2014; Jiang and Li, 2016). For matched or enriched challenge sets, the transport requirements of S5.3 apply.

This supplement states the sampling design and the external-validation requirements. The identification conditions for Equations 15 and 16, the off-policy estimators, and the transport weights required when the challenge population differs from $P_G$ are given in Supplementary Methods S5. Equation 16 compares the policies under the same candidate distribution, definition of the endpoint and context. It does not measure the total effect of a change in the generation of the candidates, in the exposure or in the reference outcome. Such changes require candidate or outcome distributions which depend on the policy, and therefore a separate estimand. Thus, fixing the distribution allows the value of the decision policy to be assessed independently of generative feedback effects.

Proxy-outcome requirements are given in Supplementary Methods S5.5. We weight explicitly, in the decision loss, the consequences of a severe false negative. Calibration, abstention efficiency, assay-selection utility, false-positive exclusions, delay, cost and animal use are secondary outcomes or components of the loss.

A credible evaluation includes prospectively generated crossings, an external validation in the archive of another sponsor or in another family of generators, and a replication in a second safety domain. We report the effect sizes and the uncertainty relative to the prespecified minimum meaningful improvement.

**References cited in this supplement**